\newif\ifpreprint
\preprinttrue
\newif\ifcomment
\newif\ifdraft
\documentclass[ALICE,manyauthors]{cernphprep}
\usepackage[comma,square,numbers,sort&compress]{natbib}
\usepackage[utf8]{inputenc}
\usepackage[T1]{fontenc}
\usepackage{hyperref}
\usepackage{lineno}
\usepackage{epsfig}
\usepackage{color}
\usepackage{xspace}
\definecolor{dgreen}{cmyk}{1.,0.,1.,0.1}      % dark green
\definecolor{orange}{cmyk}{0.,0.353,1.,0.}    % orange

\newcommand{\com}[1]    {\relax}

\newcommand{\GeVc}{\ensuremath{{\rm GeV}/c}\xspace}

\def\PbPb  {\mbox{Pb--Pb}}
\def\XeXe  {\mbox{Xe--Xe}}

\newcommand{\pt}{\ensuremath{p_{\rm T}}\xspace}
\newcommand{\mpt}{\ensuremath{[\pt]}\xspace}
\newcommand{\vn}{\ensuremath{v_{\rm n}}\xspace}

\newcommand{\vnn}[1]{\ensuremath{v_{#1}}\xspace}
\newcommand{\vnnsq}[1]{\ensuremath{\vnn{#1}^{2}}\xspace}

\graphicspath{{.img/}}

\begin{document}%
%%%%%%%%%%%%%%%  Title page %%%%%%%%%%%%%%%%%%%%%%%%
\PHyear{2021}
\PHnumber{232}      % required, will be obtained from PH
\PHdate{5 November}  % required, will be obtained from PH
\begin{titlepage}
\title{Characterizing the initial conditions of heavy-ion collisions at the LHC with mean transverse momentum and anisotropic flow correlations}
\ShortTitle{Correlations of $[p_{\rm T}]$ and $v_{\rm n}$ in Pb--Pb and Xe--Xe collisions}   % appears on right page headers
\Collaboration{ALICE Collaboration\thanks{See Appendix~\ref{app:collab} for the list of collaboration members}}
\ShortAuthor{ALICE Collaboration} % appears on left page headers, do not change

\begin{abstract}
Correlations between mean transverse momentum $[p_{\rm T}]$ and anisotropic flow coefficients $v_{\rm 2}$ or $v_{\rm 3}$ are measured as a function of centrality in Pb--Pb and Xe--Xe collisions at $\sqrt{s_{\rm NN}} = 5.02$~TeV and 5.44~TeV, respectively, with ALICE. In addition, the recently proposed higher-order correlation between $[p_{\rm T}]$, $v_{\rm 2}$, and $v_{\rm 3}$ is measured for the first time, which shows an anticorrelation for the presented centrality ranges. These measurements are compared with hydrodynamic calculations using IP-Glasma and $\rm T_{R}ENTo$ initial-state shapes, the former based on the Color Glass Condensate effective theory with gluon saturation, and the latter a parameterized model with nucleons as the relevant degrees of freedom. The data are better described by the IP-Glasma rather than the $\rm T_{R}ENTo$ based calculations. In particular, Trajectum and JETSCAPE predictions, both based on the $\rm T_{R}ENTo$ initial state model but with different parameter settings, fail to describe the measurements. As the correlations between $[p_{\rm T}]$ and $v_{\rm n}$ are mainly driven by the correlations of the size and the shape of the system in the initial state, these new studies pave a novel way to characterize the initial state and help pin down the uncertainty of the extracted properties of the quark--gluon plasma recreated in relativistic heavy-ion collisions.

\end{abstract}
\end{titlepage}                                                                                                                                              
\setcounter{page}{2}

%=====================INTRODUCTION===============================

%\section{Introduction}

The primary goal of ultrarelativistic heavy-ion collisions is to study the quark--gluon plasma (QGP)~\cite{Shuryak:1980tp}, a deconfined state of quarks and gluons, predicted by quantum chromodynamics (QCD) to emerge at extreme densities and temperatures. High-energy heavy-ion collisions at Relativistic Heavy Ion Collider (RHIC)~\cite{BRAHMS:2004adc, PHENIX:2004vcz, PHOBOS:2004zne, STAR:2005gfr} and the Large Hadron Collider (LHC) at CERN~\cite{Muller:2012zq, Roland:2014jsa} have yielded strong evidence that the QGP is observed in such collisions, enabling the study of its properties in the laboratory.
A key phenomenon that provides valuable information on the transport properties of the created QGP matter is the anisotropic expansion of the produced particles~\cite{Ollitrault:1992bk}. 
The final anisotropy can be quantified by a Fourier decomposition of the single particle azimuthal distribution~\cite{Voloshin:1994mz},
\begin{equation}
P(\varphi) = \frac{1}{2\pi} \left[ 1 + 2\, \sum_{\rm n=1}^{\infty} {v_{\rm n} \, \cos n (\varphi - \Psi_{\rm n})} \right].
\end{equation} 
Here, $\varphi$ is the azimuthal angle of the emitted particle, and $v_{\rm n}$ and $\Psi_{\rm n}$ are the $n$-th order flow coefficient and flow symmetry-plane angle, respectively. Systematic measurements on fluctuations and correlations of $v_{\rm n}$ coefficients and $\Psi_{\rm n}$ have been previously reported in Refs.~\cite{ALICE:2011ab, ATLAS:2012at,Chatrchyan:2013kba,Aad:2013xma,Aad:2014fla, Aad:2015lwa, ALICE:2016kpq,Adam:2016izf,Sirunyan:2017fts,ALICE:2017lyf,Acharya:2017zfg,Acharya:2018lmh,Acharya:2018ihu,Acharya:2019uia, Acharya:2020taj,Acharya:2021hpf}. Comprehensive comparisons with hydrodynamic model calculations provide critical information on the event average initial-state shape and the initial energy density distribution in the nuclear overlap region, as well as their event-by-event fluctuations. Additionally, they constrain the shear and bulk viscosity over entropy density ratios of the QGP, $\eta/s$ and $\zeta/s$, respectively~\cite{Heinz:2013th, Luzum:2013yya, Shuryak:2014zxa, Song:2017wtw}. Even at fixed final-state particle multiplicity, not only the shape but also the average size of the nuclear overlap region will, in general, fluctuate from event to event. These size fluctuations (at constant charged-particle density) lead to fluctuations of the pressure gradient and therefore affect the radial flow, thus influencing the transverse momentum ($\pt$) spectra of the produced particles. Arguably, the event-by-event fluctuations of the mean transverse momentum \mpt (the average transverse momentum of all particles in a single event) are more sensitive to the equation-of-state and $\zeta$/s than the measurements of the anisotropic flow~\cite{Giacalone:2020lbm,Schenke:2020uqq}. Despite their early success in describing $v_{\rm n}$ measurements, hydrodynamic models have been able to reproduce the observed \mpt fluctuations only recently~\cite{Bozek:2017elk, Bernhard:2019bmu,Everett:2020xug}. Notably the measurements of $v_{\rm n}$ and \mpt are used as independent experimental inputs for the Bayesian analyses~\cite{Bernhard:2019bmu,Everett:2020xug,Nijs:2020ors,JETSCAPE:2020shq}, which extract state-of-the-art information on the initial conditions and the final state properties, including the temperature dependence of $\eta/s$ and $\zeta/s$. 
 
Apart from the individual studies of $v_{\rm n}$ and \mpt, the interplay between radial and anisotropic flow was qualitatively investigated via anisotropic flow of identified hadrons~\cite{Huovinen:2001cy,Teaney:2000cw} and with event-shape engineering (ESE)~\cite{Adam:2015eta}. It was proposed that correlations between radial and anisotropic flow could be quantified via correlations between \mpt and \vn using a modified Pearson correlation coefficient~\cite{Bozek:2016yoj},
\begin{equation}
\rho(v_{\rm n}^{2}, [\pt]) = \frac{{\rm Cov} (v_{\rm n}^{2}, \mpt)}{\sqrt{{\rm Var}(v_{\rm n}^{2})} \, \sqrt{c_{\rm k}}},
\end{equation}
where ${\rm Cov}(v_{\rm n}^{2}, \mpt)$ is the covariance between $v_{\rm n}^{2}$ and \mpt, it can be calculated using a three-particle correlation following Eq. (1) in Ref.~\cite{Bozek:2016yoj}. The variance of $v_{n}^2$ fluctuations is given by ${\rm Var}(v_{\rm n}^{2})$ and can be measured by two- and four-particle cumulants, ${\rm Var}(v_{\rm n}^{2}) = v_{\rm n}\{2\}^{4} - v_{\rm n}\{4\}^4$. Dynamical transverse momentum correlations are given by $c_{\rm k}$~\cite{Voloshin:1999yf,ALICE:2014gvd, Bozek:2016yoj, STAR:2019dow}. As $\rho(v_{\rm n}^{2}, [\pt])$ (for n = 2, 3) can be qualitatively or even quantitatively reproduced by the initial state correlations~\cite{Schenke:2020uqq,Giacalone:2020byk,Giacalone:2020dln}, its measurements will provide valuable information on the overlap region's shape and size, and their correlations in the initial conditions. 
In particular, $\rho(v_{\rm 2}^{2}, \mpt)$ is found to be sensitive to the nuclear quadrupole deformation~\cite{Giacalone:2020awm}, adding a new tool to study the nuclear structure, which has been addressed systematically only at low energies so far~\cite{Moller:2015fba}. This $\rho(v_{\rm n}^{2}, \mpt)$ (for n = 2, 3, 4) observable has been measured previously~\cite{ATLAS:2019pvn}, which reported a clear dependence on charged-particle multiplicity, when selecting particles with $p_{\rm T} > 0.5$ GeV/$c$.

Recently, a new observable that probes the correlations between \mpt and two different $v_{\rm m}^{2}$ and $v_{\rm n}^{2}$ coefficients has been introduced~\cite{Bozek:2021zim}. This observable $\rho(v_{\rm m}^{2}, v_{\rm n}^{2}, \mpt)$ can be achieved by replacing three observables A, B, and C with $v_{\rm m}^{2}$, $v_{\rm n}^{2}$, and $\mpt$ in Eq. (24) from Ref.~\cite{Bozek:2021zim},
\begin{equation}
\small{ \rho(v_{\rm m}^{2}, v_{\rm n}^{2}, [\pt]) = \frac{ Cor(v_{\rm m}^2, v_{\rm n}^2, [\pt])}{\sqrt{{\rm Var}(v_{\rm m}^2)}\sqrt{{\rm Var}(v_{\rm n}^2)}\sqrt{c_k}} - \frac{\langle v_{\rm m}^2 \rangle}{\sqrt{{\rm Var}(v_{\rm m}^2)}} \cdot \rho_{\rm n} - \frac{\langle v_{\rm n}^2 \rangle}{\sqrt{{\rm Var}(v_{\rm n}^2)}} \cdot \rho_{\rm m} - \frac{\langle [\pt] \rangle}{\sqrt{c_{\rm k}}} \cdot \frac{SC({\rm m,n})}{\sqrt{{\rm Var}(v_{\rm m}^2)}\sqrt{{\rm Var}(v_{\rm n}^2)}}, }
\label{eq3}
\end{equation}
where $\rho_i = \rho(v_{i}^{2}, [\pt])$, $Cor(v_{\rm m}^2, v_{\rm n}^2, [\pt])$ is the correlation among $v_{\rm m}^2$, $v_{\rm n}^2$, and $\mpt$ defined in Ref.~\cite{Bozek:2021zim}, and $SC({\rm m,n})$ is the symmetric cumulant between $v_{\rm m}^2$ and $v_{\rm n}^2$~\cite{Bilandzic:2013kga, ALICE:2016kpq, ALICE:2017kwu}. The $\rho(v_{\rm m}^{2}, v_{\rm n}^{2}, [\pt])$ is constructed based on multiparticle cumulants~\cite{Bozek:2021zim,Moravcova:2020wnf}, where lower-order few particle correlations have been removed, thus, it only reflects the genuine correlations between $\mpt$, $v_{\rm n}$, and $v_{\rm m}$. It is potentially more sensitive than $\rho(v_{\rm n}^{2}, [\pt])$ to the initial conditions and is expected to be used to probe the initial momentum anisotropy~\cite{Giacalone:2020byk}, an asymmetry in the transverse pressure of the system at the interface between a pre-equilibrium description and a hydrodynamic description. The presence of an initial momentum anisotropy was predicted from first-principle considerations in the color glass condensate (CGC) effective theory of high-energy QCD~\cite{Iancu:2000hn, Ferreiro:2001qy}. However, conclusive evidence for the CGC remains elusive.

%===============================DATA SETS========================
%\section{Data sets and systematic uncertainty}

In this Letter, the measurements of the centrality dependence of correlations between flow coefficients and \mpt in Pb--Pb and Xe--Xe collisions at 5.02 TeV and 5.44 TeV, respectively, are presented. The collision centrality is determined using energy deposition in the two scintillator arrays of the V0 detector, V0A and V0C, which cover the pseudorapidity ranges of $2.8 < \eta < 5.1$ and $-3.7 < \eta <-1.7$, respectively~\cite{ALICE:2013axi,ALICE:2015juo}. Events that pass central, semi-central, or minimum-bias trigger criteria with a reconstructed primary vertex (PV) within $\pm10$ cm of the nominal interaction point along the beam direction are used. Background events are removed using information from multiple detectors as described in Ref.~\cite{ALICE:2014sbx}. A total of 245 million \PbPb\ collisions and 1.2 million \XeXe\ collisions pass these criteria. The charged-particle tracks are reconstructed using the Inner Tracking System (ITS)~\cite{Aamodt:2010aa} and the Time Projection Chamber (TPC)~\cite{Alme:2010ke}. To select only high-quality reconstructed tracks for the analysis, they are required to be in the kinematic range $0.2 < p_{\rm T} < 3.0$ GeV/$c$ and $|\eta| < 0.8$, to have more than 70 TPC space points (out of a maximum of 159), and a $\chi^2$ per degree of freedom of the track fit to the TPC space points to be lower than 2. In order to reduce the contamination from secondary particles, the distance-of-closest-approach (DCA) of the tracks to the PV must be within 2 cm in the longitudinal direction and a $p_{\rm T}$-dependent distance selection in the transverse plane, ranging from 0.2 cm at $p_{\rm T} = 0.2$ GeV/$c$ to 0.02 cm at $p_{\rm T} = 3.0$ GeV/$c$, is applied. To suppress the non-flow contaminations, which are the azimuthal angle correlations not associated to $\Psi_{\rm n}$, multiparticle correlations with the subevent method~\cite{Moravcova:2020wnf,Huo:2017nms} are applied. For this, the pseudorapidity acceptance of the central barrel is divided into three regions (subevents), A, B, and C, corresponding to $-0.8 < \eta_{\rm A} < -0.4$, $|\eta_{\rm B}| < 0.4$, and $0.4 < \eta_{\rm C} < 0.8$, respectively. Here subevents A and C are used in the $v_{\rm n}$ measurements, while subevent B is used for \mpt measurements. The $v_{\rm n}$, \mpt, and their correlations are measured in each event and corrected for detector acceptance and track-reconstruction efficiency using the latest developments based on generic framework with 3-subevent method~\cite{Bilandzic:2013kga, Huo:2017nms} including \mpt calculations.

Systematic uncertainties are estimated by varying event and track selection criteria. Uncertainties related to the selection of the event include the variation of the accepted vertex position along the beam line (9, 7 and 5 cm), and consideration of different magnetic field directions, which are analyzed separately. The resulting systematic uncertainty was found to be within 2\%. The systematic uncertainties related to track selection criteria are estimated by considering different track reconstruction algorithms and reconstruction qualities. The variations of the maximum allowed DCA to the primary vertex along the beam line and in the transverse plane result in differences less than 1\%. The quality of reconstructed tracks is varied by increasing the minimum number of space points in the TPC associated with the reconstructed track to 80 and 90, which leads to a negligible effect on the measured correlations. It is also found that the tracking efficiency exhibits a slight centrality dependence, varying by about 4\% with multiplicity. To account for this, the tracking efficiency is varied by $\pm 4\%$ in a \pt-dependent way, so that the efficiency is at its nominal value at \pt = 0.2 \GeVc and 4\% larger or smaller at \pt = 3.0 \GeVc. This yields a systematic uncertainty below 1\%. The residual non-flow contaminations are studied with the 3-subevent method~\cite{Huo:2017nms,Zhang:2021phk} and the effects are found to be negligible, which agrees with the findings from model studies~\cite{Zhang:2021phk}. Only the sources of systematic uncertainty found to be statistically significant by more than 1$\sigma$ following the procedure introduced in Ref.~\cite{Barlow:2002yb} are added in quadrature to obtain the total systematic uncertainty.

%================================RESULTS==========================

%\section{Results}

\begin{figure}[t!]
\begin{center}
\includegraphics[width=0.95\linewidth]{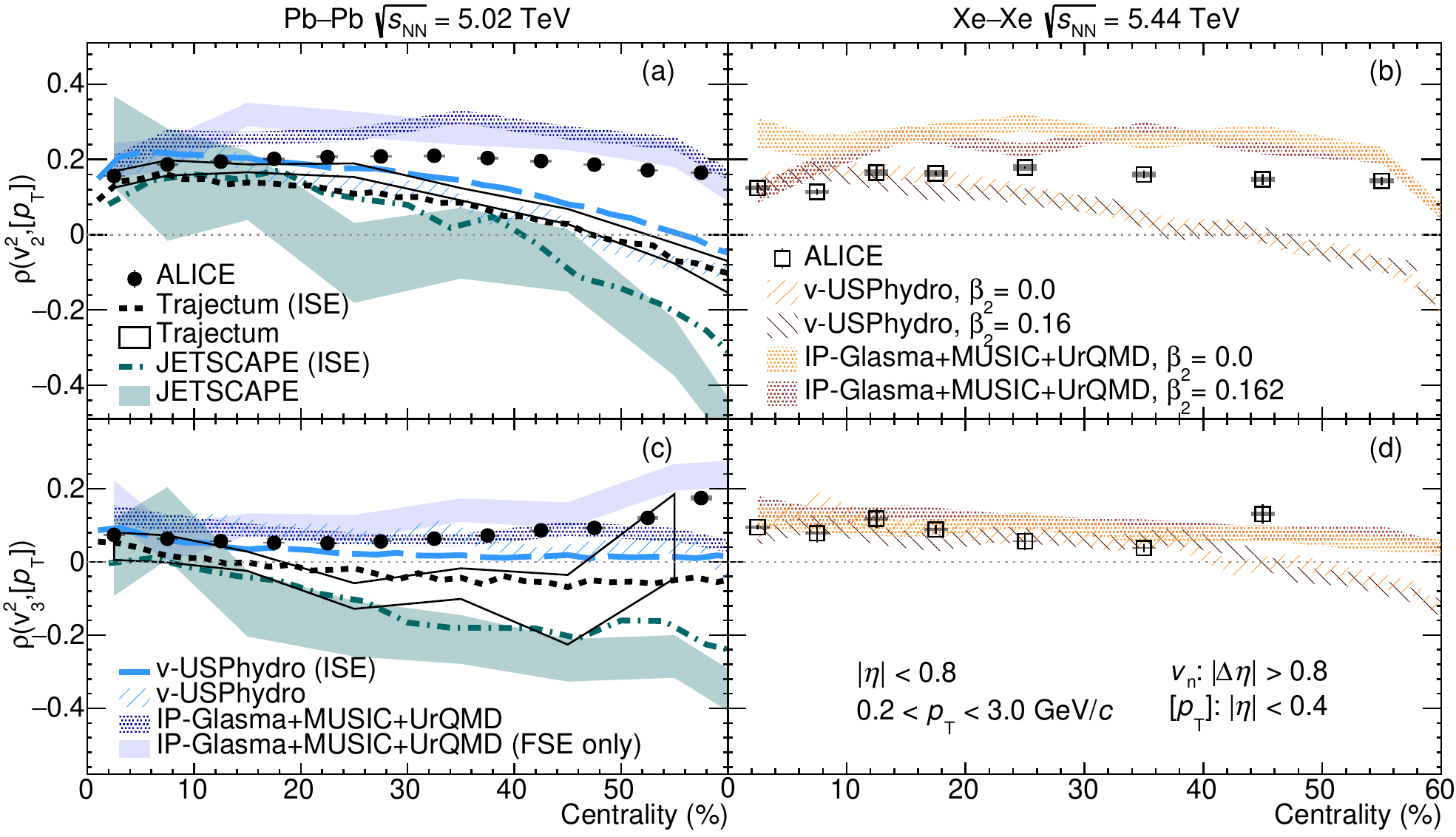}
\caption{Centrality dependence of $\rho \left(v_{\rm 2}^{2}, [ p_{\rm T} ] \right)$ (top) and $\rho \left(v_{\rm 3}^{2}, [ p_{\rm T} ] \right)$ (bottom) in Pb--Pb collisions at $\sqrt{s_{_{\rm NN}}} = 5.02$~TeV (a,c) and Xe--Xe collisions at $\sqrt{s_{_{\rm NN}}} = 5.44$~TeV (b, d). The statistical (systematic) uncertainties are shown as vertical bars (filled boxes). The initial-state estimations (ISE) are represented by lines, while IP-Glasma+MUSIC+UrQMD~\cite{Schenke:2020uqq},  v-USPhydro~\cite{Giacalone:2020dln}, Trajectum~\cite{Nijs:2020ors}, and JETSCAPE~\cite{JETSCAPE:2020shq} hydrodynamic model calculations are shown with hatched bands. }
\label{fig:rho_2_3}
\end{center}
\end{figure}

The measurements of $\rho(\vnnsq{2}, \mpt)$ and $\rho(\vnnsq{3}, \mpt)$ as a function of collision centrality in Pb--Pb collisions at $\sqrt{s_{\rm NN}} = 5.02$~TeV and Xe--Xe collisions at $\sqrt{s_{\rm NN}} = 5.44$~TeV are presented in Fig.~\ref{fig:rho_2_3}. In Pb--Pb and Xe--Xe collisions, $\rho(v_{\rm 2}^{2}, [\pt])$ has a weak centrality dependence and is positive in the considered centrality range. This implies that $v_2$ and \mpt are positively correlated and confirms the previous studies using the ESE approach~\cite{Adam:2015eta}. The presented ALICE results also qualitatively agree with the previous ATLAS measurements~\cite{ATLAS:2019pvn}. The stronger centrality dependence reported by the ATLAS Collaboration might be attributed to the different kinematic selection criteria and the choice of centrality determination. 
Hydrodynamic model calculations from the v-USPhydro model~\cite{Giacalone:2020lbm}, the Trajectum Bayesian analysis~\cite{Nijs:2020ors}, the JETSCAPE Bayesian analysis~\cite{JETSCAPE:2020shq}, and the IP-Glasma+MUSIC+UrQMD model~\cite{Schenke:2020uqq}, whenever available, are compared to data. The width of the bands illustrated in Fig.~\ref{fig:rho_2_3} denotes the statistical uncertainty of the model calculations using maximum a posteriori (MAP) parametrization, while any potential systematic uncertainty arising from different parametrizations of the models has not been evaluated. The v-USPhydro model uses $\rm T_{R}ENTo$ initial conditions tuned in Ref.~\cite{Bernhard:2016tnd} and evolved by the v-USPhydro hydrodynamic code~\cite{Giacalone:2020dln}. The Trajectum~\cite{Nijs:2020ors} and JETSCAPE~\cite{JETSCAPE:2020shq} predictions are also based on $\rm T_{R}ENTo$ initial conditions but tuned as described in Refs.~\cite{Nijs:2020ors} and~\cite{JETSCAPE:2020shq}, respectively. The IP-Glasma+MUSIC+UrQMD model uses IP-Glasma initial conditions~\cite{Schenke:2012hg, Schenke:2012wb} followed by the MUSIC hydrodynamic model~\cite{Schenke:2010rr}, coupled to a hadronic cascade model (UrQMD)~\cite{Bass:1998ca, Bleicher:1999xi}. In general, these models can quantitatively describe the previous measurements of particle $p_{\rm T}$ distributions and anisotropic flow~\cite{Giacalone:2017dud,McDonald:2016vlt}. As shown in Fig.~\ref{fig:rho_2_3}, the IP-Glasma+MUSIC+UrQMD calculations capture the general trend of the measured $\rho(v_{\rm 2}^{2}, [\pt])$ with a weak centrality dependence, qualitatively describe the measurements in Pb--Pb and Xe--Xe collisions, but slightly overestimates the data. The v-USPhydro and Trajectum calculations exhibit a strong centrality dependence, underestimate the data by more than 50\% for centrality above 30\%, and have an opposite sign with respect to data for centralities above 40\%. The discrepancies between the measurements and $\rm T_{R}ENTo$-based calculations become more pronounced with the JETSCAPE predictions~\cite{JETSCAPE:2020shq}, which become negative for semicentral collisions.

A recent study~\cite{Magdy:2021ocp} showed that $\rho(v_{\rm 2}^{2}, [\pt])$ is more sensitive to the initial conditions of the collisions rather than to the transport properties of the QGP. This is also supported by the good agreement between the various hydrodynamic calculations and the correlation coefficients calculated directly from the corresponding initial-state model, which are presented by solid or dashed lines in the top panel of Fig.~\ref{fig:rho_2_3}. These initial-state estimations (ISE) are calculated using the correlations of the energy of the fluid per unit rapidity at the initial time $\tau_{\rm 0}$ and initial anisotropy coefficient $\varepsilon_{\rm n}$~\cite{Giacalone:2020dln}. Thus, differences between the calculations of $\rho(v_{\rm 2}^{2}, [\pt])$ shown in Fig.\ref{fig:rho_2_3} are not primarily due to different hydrodynamic codes or treatments of hadronic interactions in these models. The different model predictions are rather caused by the difference in the initial energy density profile (geometric effect) and, potentially, a contribution from an initial momentum anisotropy, which is included in the IP-Glasma framework based on the CGC effective theory. The effect of initial momentum anisotropy is studied by comparing the hydrodynamic calculations using IP-Glasma initial conditions with and without initial momentum anisotropy (with only the final state effect labelled ``FSE'' shown by the light blue shadows). The IP-Glasma based calculations shown in Fig.~\ref{fig:rho_2_3} (a) are consistent with each other in the 0--60\% Pb--Pb collisions. Thus, the different descriptions of ALICE data from IP-Glasma and $\rm T_{R}ENTo$ based calculations are mainly driven by the initial geometric effects, which possibly originated from different values of $\omega$ parameter that determines the width of the colliding nucleon in the initial conditions~\cite{Moreland:2014oya}. The variable $\rho(v_{\rm 2}^{2}, [\pt])$ is the first observable for which such a significant difference is seen between $\rm T_{R}ENTo$ and IP-Glasma models. Moreover, it is found that $\rho(v_{\rm 2}^{2}, [\pt])$ is sensitive to the quadrupole deformation parameter $\beta_2$ of deformed nuclei, such as $^{238}$U or $^{129}$Xe~\cite{Giacalone:2019pca, Giacalone:2020awm}. Previous flow measurements in Xe--Xe collisions~\cite{Acharya:2018ihu} provided the estimation of $\beta_2 \approx 0.16$ for Xe using the data in the 0--5\% central collisions. The $\rho(v_{\rm 2}^{2}, [\pt])$ calculation with $\beta_2 =0$ and 0.162, based on IP-Glasma initial conditions in Fig.~\ref{fig:rho_2_3} (b), show a difference of 10--40\% in the most central collisions. Comparisons of presented ALICE measurements to IP-Glasma+MUSIC+UrQMD calculations with the two different $\beta_2$ values suggest that the data agrees better with the IP-Glasma+MUSIC+UrQMD model calculations using $\beta_2 =0.162$. Last but not least, the magnitude of $\rho(v_{\rm 2}^{2}, [\pt])$ measured in Pb--Pb collisions is larger than that in Xe--Xe collisions. Such a difference is predicted by both v-USPhydro and IP-Glasma+MUSIC+UrQMD calculations and is believed to be useful to discover a potential triaxial structure of $^{129}$Xe at the LHC energies~\cite{Bally:2021qys, Jia:2021tzt}. Comparisons with calculations that include different deformation scenarios could provide strong constraints on the $^{129}$Xe nuclear structure. This is highly non-trivial because it is not entirely clear at the moment if the nuclear structure at the LHC energies, where the partonic degree of freedom is relevant, should be quantified by the same set of parameters as the low energy nuclear structure studies.

Figure~\ref{fig:rho_2_3} (bottom) shows the centrality dependence of $\rho(v_{\rm 3}^{2}, [\pt])$ in the two colliding systems. The results are consistent within uncertainties for the considered centrality range. Both results show a positive value and rather a weak centrality dependence and a modest increase for centrality above 40\% in Pb--Pb collisions. The IP-Glasma+MUSIC+UrQMD calculations describe the presented ALICE measurements up to 50\% centrality for Pb--Pb collisions, at which point they diverge from data with opposite trends for the full IP-Glasma+MUSIC+UrQMD calculation and the one with FSE only. It is unclear yet whether this difference between IP-Glasma+MUSIC+UrQMD calculation and the one with FSE can be attributed solely to the initial momentum anisotropy originating from the contributions of CGC in the IP-Glasma model, as only statistical uncertainties are considered in these calculations. In addition, both calculations describe the ALICE measurements from Xe--Xe collisions. The v-USPhydro calculations are roughly compatible with data in Pb--Pb collisions and are consistent with ALICE measurements up to 40\% Xe--Xe collisions, then show a decreasing trend and become negative, which is not observed in the data. Like v-USPhydro, the Trajectum calculation is based on the $\rm T_{R}ENTo$ initial state model but with a different tuning; it is negative for centrality above 10\% and cannot describe the ALICE data in Pb--Pb collisions. It is also seen in Fig.~\ref{fig:rho_2_3} (bottom) that the $\rho(v_{\rm 3}^{2}, [\pt])$ calculations follow the trend of the initial-state estimations, which also show a significant difference and an opposite sign compared to the measurements. Such differences are further enhanced in the initial-state estimations from the JETSCAPE model, which does not consider the subnucleon structure. It is argued in a recent paper~\cite{Giacalone:2021clp} that the positive $\rho(v_{\rm n}^{2}, [\pt])$ for centrality up to 60\% shown in Fig.~\ref{fig:rho_2_3} suggests a nucleon size of order 0.4--0.5 fm. This is much smaller than the values obtained from the recent Bayesian analyses~\cite{Nijs:2020ors,JETSCAPE:2020shq}. Using $\rho(v_{\rm n}^{2}, [\pt])$ measurements to constrain the width of nucleon is particularly interesting. It enables a new possibility to crosscheck different determinations of the nucleon and subnucleon scales relevant to inelastic processes at high energy in different experiments, e.g., in electron--ion collisions at the future Electron-Ion Collider.

\begin{figure}[t!]
\begin{center}
\includegraphics[width=0.55\linewidth]{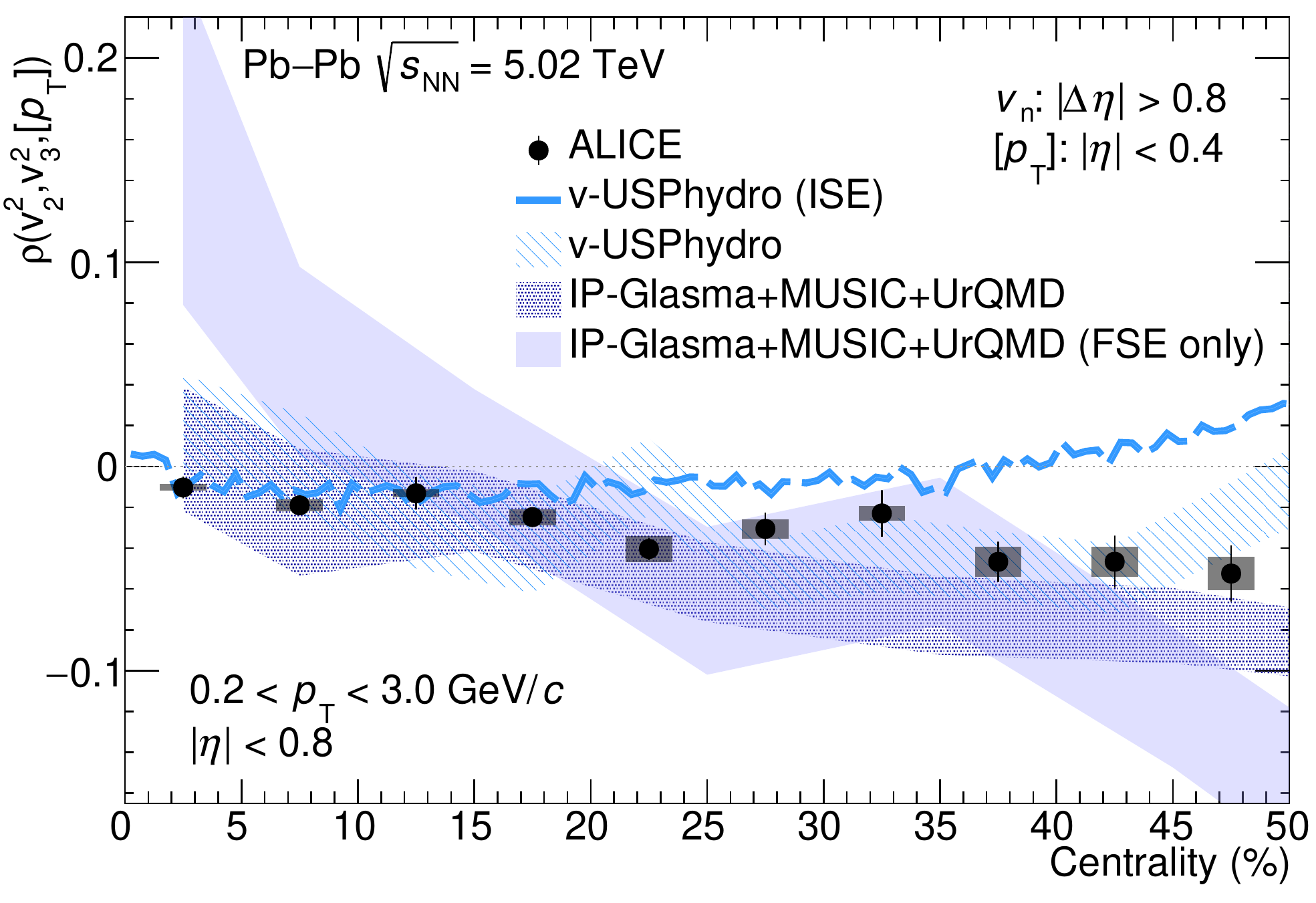}
\caption{Centrality dependence of $\rho \left(v_{\rm 2}^{2}, v_{\rm 3}^{2}, [ p_{\rm T} ] \right)$ in Pb--Pb collisions, shown by the solid circles. The statistical (systematic) uncertainties are shown as vertical bars (boxes). The initial-state estimations (ISE) as well as IP-Glasma+MUSIC+UrQMD~\cite{Schenke:2020uqq} and v-USPhydro~\cite{Giacalone:2020dln} hydrodynamic model calculations are compared with data. }
\label{fig:rho23} 
\end{center}
\end{figure}

In addition to $\rho(v_{\rm n}^{2}, [\pt])$, a newly proposed higher-order correlation involving two flow coefficients, $\rho(v_{\rm n}^{2}, v_{\rm m}^{2}, [\pt])$ as defined in Eq.~(\ref{eq3}), is expected to bring further constraints on the initial-state shape~\cite{Bozek:2021zim}. Figure~\ref{fig:rho23} reports the first measurement of $\rho(v_{\rm 2}^{2}, v_{\rm 3}^{2}, \mpt)$ as a function of centrality up to 50\% in Pb--Pb collisions at $\sqrt{s_{\rm NN}} = 5.02$~TeV while the measurement in Xe--Xe collision is very challenging due to limited data sample. First of all, negative results are observed with more than 5$\sigma$ significance from zero for the 0--50\% centrality interval, showing a non-trivial anticorrelation between $\mpt$, $v_2$, and $v_3$. It suggests an anticorrelation between entropy density, $\varepsilon_2$, and $\varepsilon_3$ in the initial stage. The full IP-Glasma+MUSIC+UrQMD and v-USPhydro calculations presented in Fig.~\ref{fig:rho23} show a flat trend with centrality as the ALICE data for the 0--40\% centrality range within uncertainties. 
As $\rho(v_{\rm 2}^{2}, v_{\rm 3}^{2}, \mpt)$ follows the trend of initial-state estimations and is mainly determined by the early stage~\cite{Bozek:2021zim}, the differences among theoretical model calculations should again mainly originate from the initial conditions. Due to the large uncertainties of the model calculations, one cannot clearly establish a preference from comparisons with the ALICE data. Furthermore, the calculation of IP-Glasma with only the final state effect (FSE) shows a strong centrality dependence. It is positive in central collisions and gradually becomes negative in peripheral collisions. This trend is different from that of full IP-Glasma+MUSIC+UrQMD calculations and the experimental data.

%================================SUMMARY==========================
%\section{Summary}

In summary, the centrality dependence of correlations between anisotropic flow coefficients and mean transverse momentum in Pb--Pb and Xe--Xe collisions has been presented. These first ALICE measurements of $\rho(v_{\rm 2}^{2}, \mpt)$ and $\rho(v_{\rm 3}^{2}, \mpt)$ are found to be positive in both collision systems for the presented centrality ranges, confirming the positive correlations between $v_2$ (or $v_{\rm 3}$) and $\mpt$. 
In addition, the higher-order correlation coefficient $\rho(v_{\rm 2}^{2}, v_{\rm 3}^{2}, \mpt)$ has been measured and found to be negative in Pb--Pb collisions, suggesting an anticorrelation between entropy density, initial anisotropy coefficients $\varepsilon_2$ and $\varepsilon_3$ in the initial stage. The experimental data reported in this Letter can be described by hydrodynamic models using IP-Glasma initial conditions, while they are not well reproduced by calculations using different tunings of the $\rm T_{R}ENTo$ initial conditions. Such discrepancies cannot be attributed to the effect of the initial momentum anisotropy predicted by the CGC framework, as its impact is insignificant in the presented centrality ranges. Instead, the discrepancies are expected to arise from different geometric effects in the initial state. It should also be emphasized that the state-of-the-art extraction of the QGP's transport coefficients relies on Bayesian analyses, which are all based on the $\rm T_{R}ENTo$ model so far. The presented discrepancy between ALICE data and $\rm T_{R}ENTo$ model-based calculations requires in-depth investigations, and the impact on the extraction of the properties of QGP should be explored further. The presented ALICE measurements are crucial on this matter, and including the presented data in the Bayesian global fitting could be a valuable step toward a better constraint on the initial state in nuclear collisions. Last but not least, the presented measurements in Xe--Xe collisions open a new window to study the nuclear structure using ultrarelativistic heavy-ion collisions at the LHC.

%%%%%%%%%%%%%%%%%%%%%%%%%%%%%%%%%%%%%%%%%%%%%%%%%%
\newenvironment{acknowledgement}{\relax}{\relax}
%%%%%%%%%%%%%%%%%%%%%%%%%%%%%%%%%%%%%%%%%%%%%%%%%%
\begin{acknowledgement}
\section*{Acknowledgments}
The ALICE Collaboration would like to thank Giuliano Giacalone (v-USPhydro), Chun Shen, Bjoern Schenke (IP-Glasma), Wilke van der Schee (Trajectum), Julia Velkovska, Andi Mankolli (JETSCAPE), Piotr Bozek (Glauber+MUSIC) for providing the latest predictions from the state-of-the-art models.
% Version: 2021-10-27

The ALICE Collaboration would like to thank all its engineers and technicians for their invaluable contributions to the construction of the experiment and the CERN accelerator teams for the outstanding performance of the LHC complex.
The ALICE Collaboration gratefully acknowledges the resources and support provided by all Grid centres and the Worldwide LHC Computing Grid (WLCG) collaboration.
The ALICE Collaboration acknowledges the following funding agencies for their support in building and running the ALICE detector:
A. I. Alikhanyan National Science Laboratory (Yerevan Physics Institute) Foundation (ANSL), State Committee of Science and World Federation of Scientists (WFS), Armenia;
Austrian Academy of Sciences, Austrian Science Fund (FWF): [M 2467-N36] and Nationalstiftung f\"{u}r Forschung, Technologie und Entwicklung, Austria;
Ministry of Communications and High Technologies, National Nuclear Research Center, Azerbaijan;
Conselho Nacional de Desenvolvimento Cient\'{\i}fico e Tecnol\'{o}gico (CNPq), Financiadora de Estudos e Projetos (Finep), Funda\c{c}\~{a}o de Amparo \`{a} Pesquisa do Estado de S\~{a}o Paulo (FAPESP) and Universidade Federal do Rio Grande do Sul (UFRGS), Brazil;
Ministry of Education of China (MOEC) , Ministry of Science \& Technology of China (MSTC) and National Natural Science Foundation of China (NSFC), China;
Ministry of Science and Education and Croatian Science Foundation, Croatia;
Centro de Aplicaciones Tecnol\'{o}gicas y Desarrollo Nuclear (CEADEN), Cubaenerg\'{\i}a, Cuba;
Ministry of Education, Youth and Sports of the Czech Republic, Czech Republic;
The Danish Council for Independent Research | Natural Sciences, the VILLUM FONDEN and Danish National Research Foundation (DNRF), Denmark;
Helsinki Institute of Physics (HIP), Finland;
Commissariat \`{a} l'Energie Atomique (CEA) and Institut National de Physique Nucl\'{e}aire et de Physique des Particules (IN2P3) and Centre National de la Recherche Scientifique (CNRS), France;
Bundesministerium f\"{u}r Bildung und Forschung (BMBF) and GSI Helmholtzzentrum f\"{u}r Schwerionenforschung GmbH, Germany;
General Secretariat for Research and Technology, Ministry of Education, Research and Religions, Greece;
National Research, Development and Innovation Office, Hungary;
Department of Atomic Energy Government of India (DAE), Department of Science and Technology, Government of India (DST), University Grants Commission, Government of India (UGC) and Council of Scientific and Industrial Research (CSIR), India;
Indonesian Institute of Science, Indonesia;
Istituto Nazionale di Fisica Nucleare (INFN), Italy;
Japanese Ministry of Education, Culture, Sports, Science and Technology (MEXT), Japan Society for the Promotion of Science (JSPS) KAKENHI and Japanese Ministry of Education, Culture, Sports, Science and Technology (MEXT)of Applied Science (IIST), Japan;
Consejo Nacional de Ciencia (CONACYT) y Tecnolog\'{i}a, through Fondo de Cooperaci\'{o}n Internacional en Ciencia y Tecnolog\'{i}a (FONCICYT) and Direcci\'{o}n General de Asuntos del Personal Academico (DGAPA), Mexico;
Nederlandse Organisatie voor Wetenschappelijk Onderzoek (NWO), Netherlands;
The Research Council of Norway, Norway;
Commission on Science and Technology for Sustainable Development in the South (COMSATS), Pakistan;
Pontificia Universidad Cat\'{o}lica del Per\'{u}, Peru;
Ministry of Education and Science, National Science Centre and WUT ID-UB, Poland;
Korea Institute of Science and Technology Information and National Research Foundation of Korea (NRF), Republic of Korea;
Ministry of Education and Scientific Research, Institute of Atomic Physics, Ministry of Research and Innovation and Institute of Atomic Physics and University Politehnica of Bucharest, Romania;
Joint Institute for Nuclear Research (JINR), Ministry of Education and Science of the Russian Federation, National Research Centre Kurchatov Institute, Russian Science Foundation and Russian Foundation for Basic Research, Russia;
Ministry of Education, Science, Research and Sport of the Slovak Republic, Slovakia;
National Research Foundation of South Africa, South Africa;
Swedish Research Council (VR) and Knut \& Alice Wallenberg Foundation (KAW), Sweden;
European Organization for Nuclear Research, Switzerland;
Suranaree University of Technology (SUT), National Science and Technology Development Agency (NSDTA) and Office of the Higher Education Commission under NRU project of Thailand, Thailand;
Turkish Energy, Nuclear and Mineral Research Agency (TENMAK), Turkey;
National Academy of  Sciences of Ukraine, Ukraine;
Science and Technology Facilities Council (STFC), United Kingdom;
National Science Foundation of the United States of America (NSF) and United States Department of Energy, Office of Nuclear Physics (DOE NP), United States of America.

\end{acknowledgement}

%%%%%%%%%%%%%%%%%%%%%%%%%%%%%%%%%%%%%%%%%%%%%%%%%%
\bibliographystyle{utphys}
\bibliography{bibliography}
%%%%%%%%%%%%%%%%%%%%%%%%%%%%%%%%%%%%%%%%%%%%%%%%%%
\newpage
\appendix
%%%%%%%%%%%%%%%%%%%%%%%%%%%%%%%%%%%%%%%%%%%%%%%%%%
\section{The ALICE Collaboration}
\label{app:collab}
%%%%%%%%%%%%%%%%%%%%%%%%%%%%%%%%%%%%%%%%%%%%%%%%%%

 \begin{flushleft}

\bigskip 

S.~Acharya$^{\rm 143}$, 
D.~Adamov\'{a}$^{\rm 97}$, 
A.~Adler$^{\rm 75}$, 
J.~Adolfsson$^{\rm 82}$, 
G.~Aglieri Rinella$^{\rm 34}$, 
M.~Agnello$^{\rm 30}$, 
N.~Agrawal$^{\rm 54}$, 
Z.~Ahammed$^{\rm 143}$, 
S.~Ahmad$^{\rm 16}$, 
S.U.~Ahn$^{\rm 77}$
% ...

\bigskip

\bigskip 

\textbf{\Large Affiliation Notes}

\bigskip 

% ...
$^{\rm VII}$ Also at: University of Kansas, Lawrence, Kansas, United States\\

\bigskip

\bigskip 

\textbf{\Large Collaboration Institutes}

\bigskip 

% ...
$^{16}$ Department of Physics, Aligarh Muslim University, Aligarh, India\\
% ...
$^{30}$ Dipartimento DISAT del Politecnico and Sezione INFN, Turin, Italy\\
% ...
$^{34}$ European Organization for Nuclear Research (CERN), Geneva, Switzerland\\
% ...
$^{54}$ INFN, Sezione di Bologna, Bologna, Italy\\
% ...
$^{75}$ Johann-Wolfgang-Goethe Universit\"{a}t Frankfurt Institut f\"{u}r Informatik, Fachbereich Informatik und Mathematik, Frankfurt, Germany\\
% ...
$^{77}$ Korea Institute of Science and Technology Information, Daejeon, Republic of Korea\\
% ...
$^{82}$ Lund University Department of Physics, Division of Particle Physics, Lund, Sweden\\
% ...
$^{97}$ Nuclear Physics Institute of the Czech Academy of Sciences, \v{R}e\v{z} u Prahy, Czech Republic\\
% ...
$^{143}$ Variable Energy Cyclotron Centre, Homi Bhabha National Institute, Kolkata, India\\
% ...

\bigskip 

\end{flushleft}

%============================

\end{document}